\documentclass[journal=jcisd8,manuscript=article]{achemso}

\usepackage[version=3]{mhchem} 
\usepackage{xcolor}

\author{Thiago Buzelli}
\affiliation[UFABC]
{Center for Natural and Human Sciences (CCNH) of Federal University of ABC (UFABC)}
\author{Bruno Ipaves}
\email{ipaves.bruno@ufabc.edu.br}
\affiliation[UFABC]
{Center for Natural and Human Sciences (CCNH) of Federal University of ABC (UFABC)}
\author{Wanda Pereira Almeida}
\affiliation[UNICAMP]
{Institute of Chemistry, State University of Campinas (UNICAMP)}
\author{Douglas Soares Galvao}
\email{galvao@ifi.unicamp.br}
\affiliation[UNICAMP]
{Applied Physics Department and Center for Computational Engineering and Sciences, State University of Campinas (UNICAMP)}
\author{Pedro Alves da Silva Autreto}
\email{pedro.autreto@ufabc.edu.br}
\affiliation[UFABC]
{Center for Natural and Human Sciences (CCNH) of Federal University of ABC (UFABC)}

\title[An \textsf{achemso} demo]
  {Machine Learning-based Analysis of Electronic Properties as Predictors of Anticholinesterase Activity in Chalcone Derivatives\footnote{A footnote for the title}}

\abbreviations{IR,NMR,UV}

\begin{document}







\begin{abstract}
 In this study, we investigated the correlation between the electronic properties of anticholinesterase compounds and their biological activity. While the methodology of such correlation is well-established and has been effectively utilized in previous studies, we employed a more sophisticated approach: machine learning. Initially, we focused on a set of $22$ molecules sharing a common chalcone skeleton and categorized them into two groups based on their IC$_{50}$ indices: active and inactive. Utilizing the open-source software Orca, we conducted calculations to determine the geometries and electronic structures of these molecules. Over a hundred parameters were collected from these calculations, serving as the foundation for the features used in machine learning. These parameters included the Mulliken and Lowdin electronic populations of each atom within the skeleton, molecular orbital energies, and Mayer's free valences. Through our analysis, we developed numerous models and identified several successful candidates for effectively distinguishing between the two groups. Notably, the most informative descriptor for this separation relied solely on electronic populations and orbital energies. By understanding which computationally calculated properties are most relevant to specific biological activities, we can significantly enhance the efficiency of drug development processes, saving both time and resources.
\end{abstract}

\section{Introduction}
Developing novel pharmaceutical compounds is a complex, resource-intensive, and time-consuming process. While drugs promise to improve life expectancy and deliver effective disease treatment, they also pose the inherent risk of adverse reactions and use abuse. The pipeline development of new pharmacological products generally involves multi-sequential stages, such as: 1. identification and discovery of compounds exhibiting therapeutic activity; 2. rigorous in vitro testing to assess their biological properties; 3. comprehensive in vivo studies to investigate their dynamics in animal models; 4. human clinical trials \cite{lombardino2004role}.

Optimizing the selection of new active compounds through computer simulations offers significant appeal for cost reduction, reagent optimization, and accelerated product development. Computational approaches, such as docking, allow the analyses of molecule-ligand interactions to identify suitable candidates that fit (energetically and geometrically) into protein pocket binding sites \cite{azam2013molecular}. In principle, selecting structures with a good potential for a desired pharmaceutical activity through interaction energy values is effectively possible. Other approaches aim to correlate geometric features, chemical composition, and electronic structure data with biological activity.

The use of computer-generated models has already achieved remarkable success in correlating biological activity with molecular electronic features obtained from electronic structure calculations. For instance, the electronic structure of acids can be directly linked to the biological activity of their derivatives \cite{NOVAK2010242}. Additionally, there is a correlation between spectroscopy data and the biological activity of steroids, highlighting the significance of the molecules' skeletal composition  \cite{NOVAK1999233}. Other studies established a correlation between polycyclic aromatic hydrocarbons (PAHs) electronic structure and their carcinogenic potential through the so-called universal quantum-molecular descriptors \cite{braga2003structure}. Recently, Autreto and co-authors successfully established a correlation between the electronic indices and biological properties of novel febrifugine derivatives \cite{autreto2008febrifugine}. Finally, the use of the electronic theory of cancer (ETC) revealed significant correlations between the PAHs' carcinogenicity and their energy difference values between the highest occupied molecular orbital (HOMO) and the second-highest occupied molecular orbital (HOMO$_{-1}$), as well as with localized charge density within some molecular regions  \cite{barone1996theoretical, braga1999identifying}.

Recent advances in machine learning (ML) techniques have been successfully used for complex problems characterized by large volumes of dynamic, unstructured, and error-containing data. With the exponential growth of generating data and rapid technological advancements in software and computer hardware, we entered the era of big data science \cite{Schleder_2019}. Data science combines mathematical principles, computer science, statistical analyses, and programming techniques to address multifaceted problems. By exploiting the computational power of modern computers and advanced complex algorithms, ML can create a large number of models from a single data set. This capability allows us to compare and combine models, thereby enhancing the quality of the analyses in extracting and identifying correlations even with sparse data \cite{Schleder_2019}. It is worth emphasizing that the use of ML extends beyond problems involving large data sets. In fact, it can be instrumental in elucidating relationships or even unveiling underlying laws governing complex phenomena where scientific knowledge remains undiscovered or is too intricate to comprehend through traditional approaches \cite{butler2018machine}.

In this study, we carried out an investigation on a set of $22$ molecules categorized as anticholinesterases. These compounds exhibit the ability to inhibit different forms of cholinesterases and have found applications in diverse domains such as pesticide development, glaucoma treatment, and myasthenia management \cite{sakata2017effect}. Notably, anticholinesterases have received considerable attention in the context of Alzheimer's therapy, with three out of the four drugs currently used for its treatment belonging to the class of anticholinesterase agents (namely, galantamine, donepezil, and rivastigmine) \cite{sakata2017effect}. 

In the present work, We have developed a new approach to identifying active and inactive molecules by integrating ML techniques with electronic structure parameters. We have obtained significantly enhanced efficiency in determining the prediction limits of our models through an extensive analysis of hundreds of thousands of combinations simultaneously. Based on that, we developed an optimized computational strategy to identify active and/or inactive molecules with specific biological activity.

\section{Methodology}

\subsection{Target molecules}
\begin{figure}
    \centering
    \includegraphics[height=5cm]{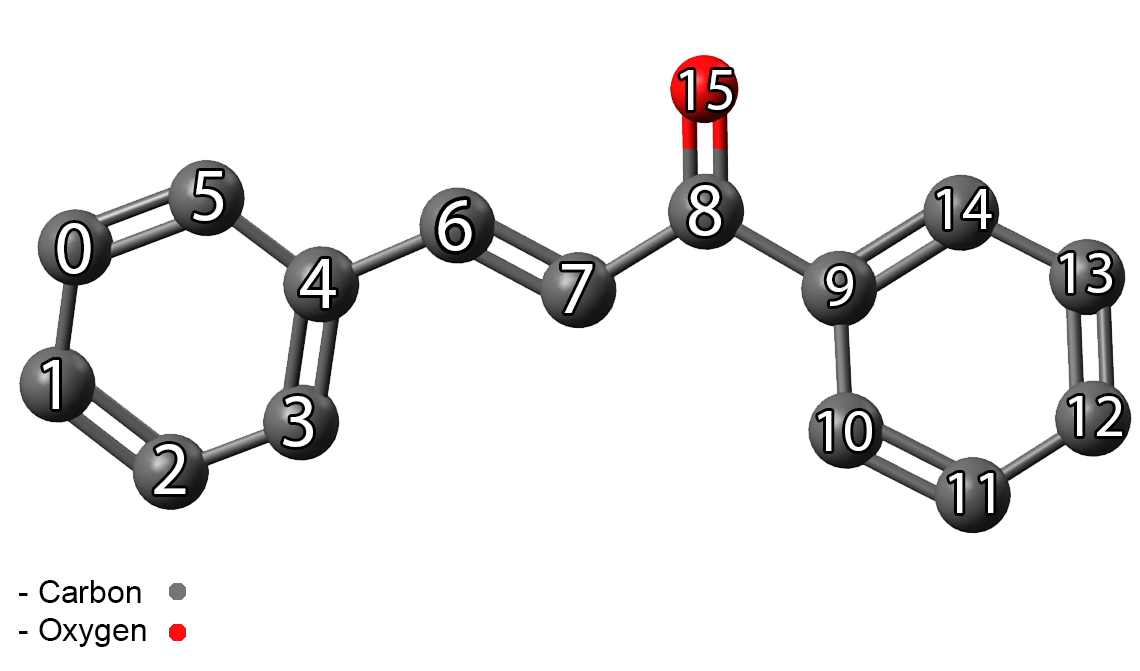}
    \caption{Skeletal structure (chalcone group) of our anticholinesterase group set. Each atom within the skeleton represents a point where multiple electronic parameters are attributed.}
    \label{fig:skeletal-structure}
\end{figure}

\begin{figure}[p]
\centering
  \includegraphics[height=18cm]{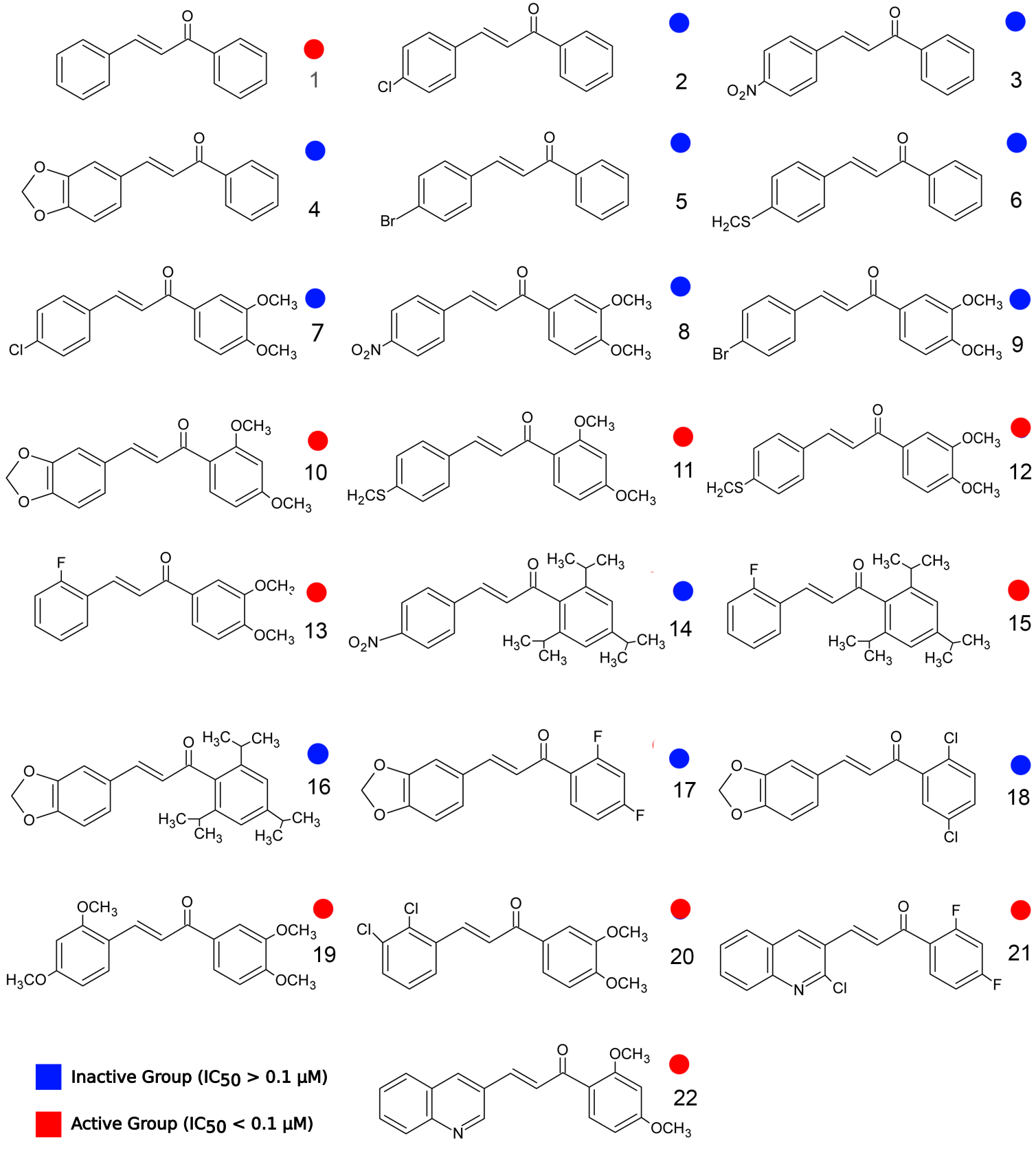}
  \caption{Molecules labeled as active are represented by a red circle on their right side, while molecules labeled as inactive are represented by a blue one on their right side.}
  \label{fgr:molecules}
\end{figure}

The molecules under investigation exhibit a shared structural framework, as can be seen from Figures \ref{fig:skeletal-structure} and \ref{fgr:molecules}. This characteristic is important as it makes it easier to obtain structural correlations across the group set. The biological activity of these molecules is presented based on the IC$_{50}$ index of each molecule in Table \ref{tbl:categorization}.

To determine the IC$_{50}$ values, the compounds were diluted in phosphate buffer (pH 7.4) and DMSO at six different concentrations: 200, 150, 100, 50, 25, and 10 micromolar. These diluted solutions were then tested against acetylcholinesterase. The acetylcholinesterase enzyme used in the study was obtained from Sigma-Aldrich Co and was sourced from Electrophorus electricus (electric eel), Type VI S. The assay was conducted using Ellman's colorimetric method. It should be noted that the maximum concentration of DMSO used in the experiments was 1\% \cite{ELLMAN196188, sakata2017effect}.

\begin{table}[hbt]
\small
\centering
  \caption{\ Categorization of molecules into two distinct groups (active and inactive) according to their IC$_{50}$ values.}
  \label{tbl:categorization}
   \begin{tabular}{lll}
    \hline
   ID & IC$_{50}$ & Group \\
    \hline
1  & 0.039  & Active \\
2  & 1.471  & Inactive \\
3  & 1.480  & Inactive \\
4  & 0.261  & Active \\
5  & 2.567  & Inactive \\
6  & 4.813  & Inactive \\
7  & 0.172  & Inactive \\
8  & 0.230  & Inactive \\
9  & 0.170  & Inactive \\
10 & 0.050  & Active \\
11 & 0.020  & Active \\
12 & 0.031  & Active \\
13 & 0.018  & Active \\
14 & 0.528  & Inactive \\
15 & 0.028  & Active \\
16 & 0.739  & Inactive \\
17 & 1.023  & Inactive \\
18 & 2.528  & Inactive \\
19 & 0.008  & Active \\
20 & 0.026  & Active \\
21 & 0.035  & Active \\
22 & 0.027  & Active \\
    \hline
    \end{tabular}
\end{table}

Based on the IC$_{50}$ values, the group of molecules was divided into two subgroups: active and inactive, as indicated in Figure \ref{fgr:molecules}. The IC$_{50}$ index represents the potency of the inhibitor compound in inhibiting 50\% of a biological or biochemical process. 
A low IC$_{50}$ value indicates a more efficient and powerful inhibitory activity of the molecule. We considered molecules with an IC$_{50}$ value below 0.1 µM (0.1 x $10^3$ mol/m$^3$) as belonging to the active subgroup, while those with IC$_{50}$ values above this threshold belonging to the inactive subgroup. These are typical classification values found in the literature.

\subsection{Calculation of electronic indices}

After classifying the molecules into these two distinct subgroups, the subsequent procedure involved the computation of electronic indices associated with each element of the molecular skeleton. These indices will serve as our (classification active/inactive) feature set and form the foundation for creating correlation models using machine learning algorithms. 

The electronic structure of the investigated molecules was obtained using the Orca computational code \cite{neese2012orca}. We used a semi-empirical approach based on the Hartree-Fock method \cite{fischer1977hartree, szabo2012modern, ramachandran2008computational}. The molecular geometries were optimized using the Quasi-Newton BFGS algorithm, the PM3 semi-empirical method, and a def2-SVP basis set \cite{schafer1992fully}. The SCF convergence criteria in ORCA for geometry optimizations were set to TightSCF, with an energy change of $1.0 \times 10^{-8}$ au, $\text{TolE} = 5 \times 10^{-6}$, $\text{TolRMSG} = 1 \times 10^{-4}$, $\text{TolMaxG} = 3 \times 10^{-4}$, $\text{TolRMSD} = 2 \times 10^{-3}$, and $\text{TolMaxD} = 4 \times 10^{-3}$ \cite{neese2012orca}.

\subsection{Electronic indices - features}
Using the optimized geometries of the anticholinesterase molecules (Figure \ref{fgr:molecules}), we calculated the following PM3 electronic indices for each skeleton atom:

\begin{itemize}
\item Mulliken and Lowdin orbital charges (s and p).
\item Mulliken and Lowdin atomic charges.
\item Orbital Energies values: HOMO$_{-3}$, HOMO$_{-2}$, HOMO$_{-1}$, HOMO, GAP, LUMO, LUMO$_{+1}$, LUMO$_{+2}$, LUMO$_{+3}$.
\item Mayer valency.
\end{itemize}

\noindent
HOMO, LUMO, and GAP refer to Highest Occupied Molecular Orbital, Lowest Unoccupied Molecular Orbital, LUMO-HOMO energy value, respectively.

To identify the best descriptors, we used an algorithm called SISSO (Sure Independence Screening and Sparsifying Operator), which combines two techniques: SIS (Sure Independence Screening) and OS (Sparsifying Operator). The SIS technique is related to reducing the dimensionality of the system, while the OS is the sparsifying operator used in the compressed sensing approach \cite{ouyang2018sisso}. 

The SISSO input file consists of a worksheet that contains a molecule (labeled example) on each line, collectively forming the data set. Each column associated with a labeled example represents a feature, and the collection of these columns composes the feature set. The SISSO output provides a ranking of maps indicating the most effective separation between the active and inactive data sets. This separation metric is determined by a pair of features known as descriptors. These descriptors assign a point to each labeled example, resulting in the formation of a two-dimensional graph that depends on the chosen descriptors.

The set of input data, known as the training set, serves as the basis for the learning algorithm. Prior to implementing machine learning methods, a subset of the data set, molecules in our case, is typically reserved for testing the model. This subset is referred to as the test set. In our study, we have separated molecules $8$ and $9$ from the inactive group and molecules $10$ and $11$ from the active group, which constitute our test set. The remaining molecules form our training set, which is used in the machine learning algorithm to establish a correlation model between electronic indices and biological activity.

\section{Results and discussions}

We have performed statistical analyses based on the Pearson correlation coefficient to select the most relevant features (Figure \ref{S1} in supplementary info). We executed a filtering process by excluding parameters with an absolute value of the Pearson coefficient ($|\rho|$) greater than or equal to $0.8$. This process resulted in a refined feature set consisting of $35$ attributes. We refer to this curated collection as Pearson's feature set. Here are the specific parameters that compose this set:

\begin{itemize}
    \item Orbital energies: HOMO$_{-3}$, HOMO$_{-1}$, HOMO, LUMO$_{+1}$
     \item Mulliken population analysis: (M0, M1, M2, M3, M4, M5, M6, M8, M10, M11, M12, M13, M15). MX indicates the Mulliken charge on atom X; (MS7, MS9, MS10). MYX Indicates the Mulliken charge in the Y orbital (which can be S or P) of the X atom.
     \item Lowdin population analysis: (L2, L4, L6, L7, L8, L10). LX indicates the Lowdin charge on the X atom; (LS0, LS1, LS5, LS7, LS11, LS13, LS14, LS15). LYX indicates the Lowdin charge on the Y orbital of the X atom.
     \item Mayer's free valence analysis: All attributes pertaining to this analysis were removed.
\end{itemize}

For each feature set, we generated a large number of models. However, on average, only 20 of them exhibited a correlation of 100\% accuracy with the biological activity of the molecules, as determined by the electronic indices. This correlation allowed us to effectively classify the molecules into two distinct groups: active and inactive. Consequently, we successfully mitigated underfitting problems, although the limited data size still posed a risk of overfitting.

To assess the performance of these models, we carried out a validation procedure using an independent test set. As a result, we narrowed down the selection to four models for the Pearson feature set. These models, except for orbital charges, are consistent with the discussions previously presented in the literature using other approaches \cite{barone2000electronic}. The following parameter models achieved a perfect 100\% accuracy in separating active and inactive molecules:

\begin{enumerate}
\item \begin{equation*}
\left[\frac{\textnormal{M}10/\textnormal{M}4}{\textnormal{M}1-\textnormal{M}3}\right],\left[\frac{\textnormal{M}10/\textnormal{L}4}{\textnormal{M}1-\textnormal{M}3}\right]
\end{equation*}

\item \begin{equation*}
\left[\frac{\textnormal{HOMO}_{-1}-\textnormal{HOMO}}{\textnormal{M}1-\textnormal{M}3}\right],\left[\frac{\textnormal{L}10/\textnormal{L}2}{\textnormal{M}1-\textnormal{M}3}\right]
\end{equation*}

\item \begin{equation*}
\left[\frac{\textnormal{HOMO}_{-1}-\textnormal{HOMO}}{\textnormal{M}1-\textnormal{M}3}\right],\left[\frac{\textnormal{M}10/\textnormal{M}4}{\textnormal{M}1-\textnormal{M}3}\right]
\end{equation*}

\item \begin{equation*}
\left[\frac{\textnormal{HOMO}_{-1}-\textnormal{HOMO}}{\textnormal{M}1-\textnormal{M}3}\right],\left[\frac{\textnormal{M}10/\textnormal{L}4}
{\textnormal{M}1-\textnormal{M}3}\right]
\end{equation*}
\end{enumerate} 

It is worth noting that among the four models, there is a recurring presence of factors such as (M1-M3) and ($\Delta H$ = HOMO$_{-1}$ - HOMO), indicating their significance in the separation of the molecular groups. It is important to emphasize that the term $\Delta H$ appears in numerous studies related to structure-activity correlations.

Among these models, models $1$, $2$, and $4$ incorporate a combination of Mulliken and Lowdin charges. Despite their shared purpose of describing the same physical quantities, these two charge schemes have distinct mathematical foundations. Consequently, we consider only model 3 as an appropriate approach for correlating biological activity with electronic indices. Taking into account the aforementioned observations, we can conclude that descriptor 3 yields the most favorable outcome among all the descriptors used in the machine learning methodology.

\begin{figure}[p]
\centering
\includegraphics[height=20cm]{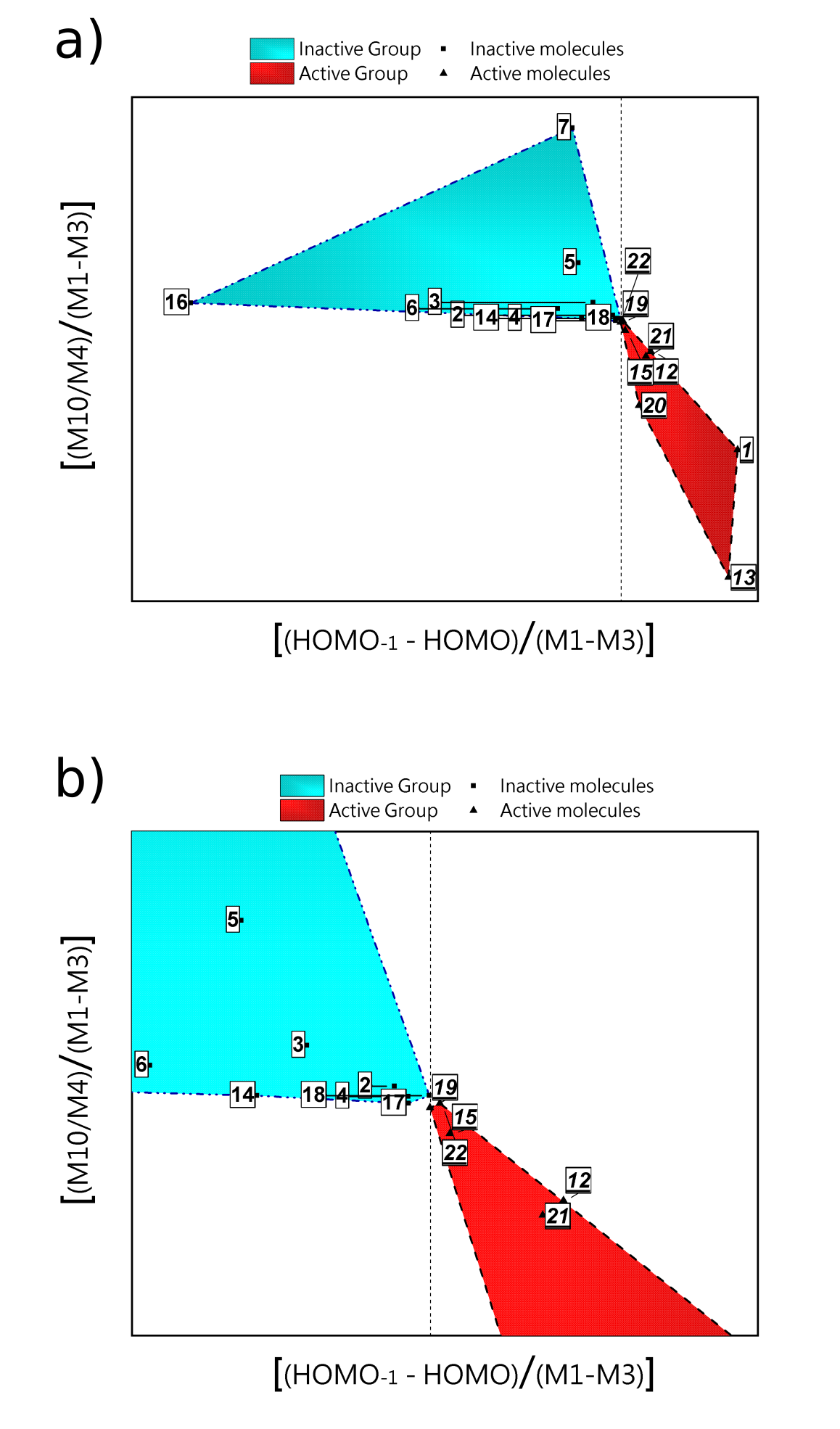}
\caption{(a) A two-dimensional map of machine learning results using descriptor 3 and the training set, and (b) a selected zoomed part of the map. The final descriptors are plotted on the axes, and it is evident that the groups do not overlap in the border region. The system metric indicates a negative overlap (separation) of -3.20.}
\label{fgr:map-descriptors-1}
\end{figure}

Figure \ref{fgr:map-descriptors-1} illustrates a two-dimensional map of the descriptors, which rely on Mulliken's charge description and the differences in orbital energy values ($\Delta H$). This model is specifically tailored for the training set, with four molecules chosen for the test set. Notably, there is a well-defined separation between the active and inactive groups, with no observed intersections.

\begin{figure}[p]
\centering
  \includegraphics[height=20cm]{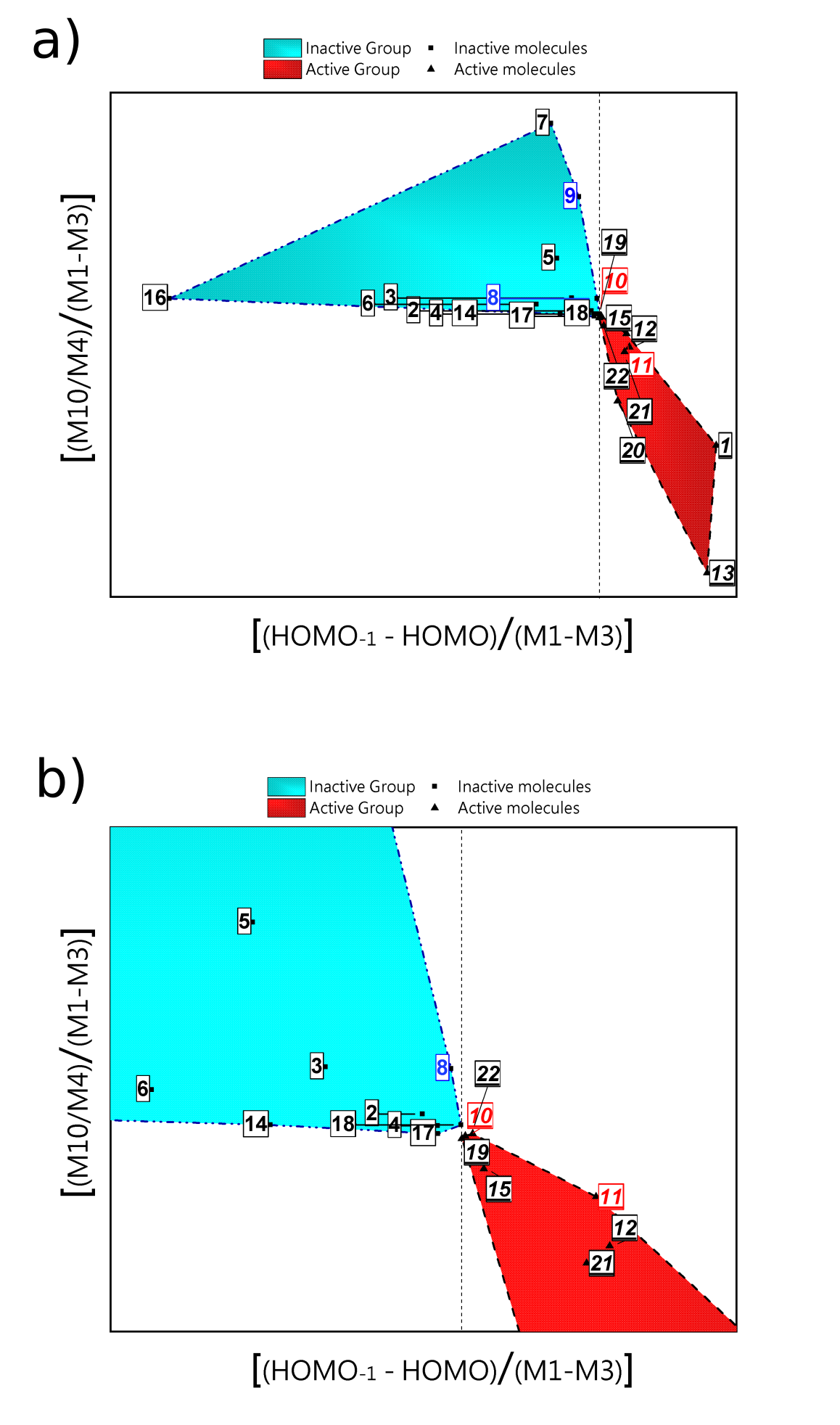}
  \caption{(a) A two-dimensional map of machine learning results, including an externally calculated test set, and (b) a selected zoomed region of the map. The figure demonstrates that the model effectively describes the groups, even though the molecules only slightly redefine the boundaries.}
  \label{fgr:map-descriptors-2}
\end{figure}

Figure \ref{fgr:map-descriptors-2} presents the ML model, based on the training set, expanded with externally calculated descriptor values for the test set molecules. These additional test set data points slightly alter the boundary definitions as they exhibit a closer alignment with their respective groups. Remarkably, it is worth emphasizing that the horizontal descriptor alone is capable of achieving 100\% accuracy in classifying the groups into active and inactive categories.

\begin{figure}[p]
\centering
  \includegraphics[height=20cm]{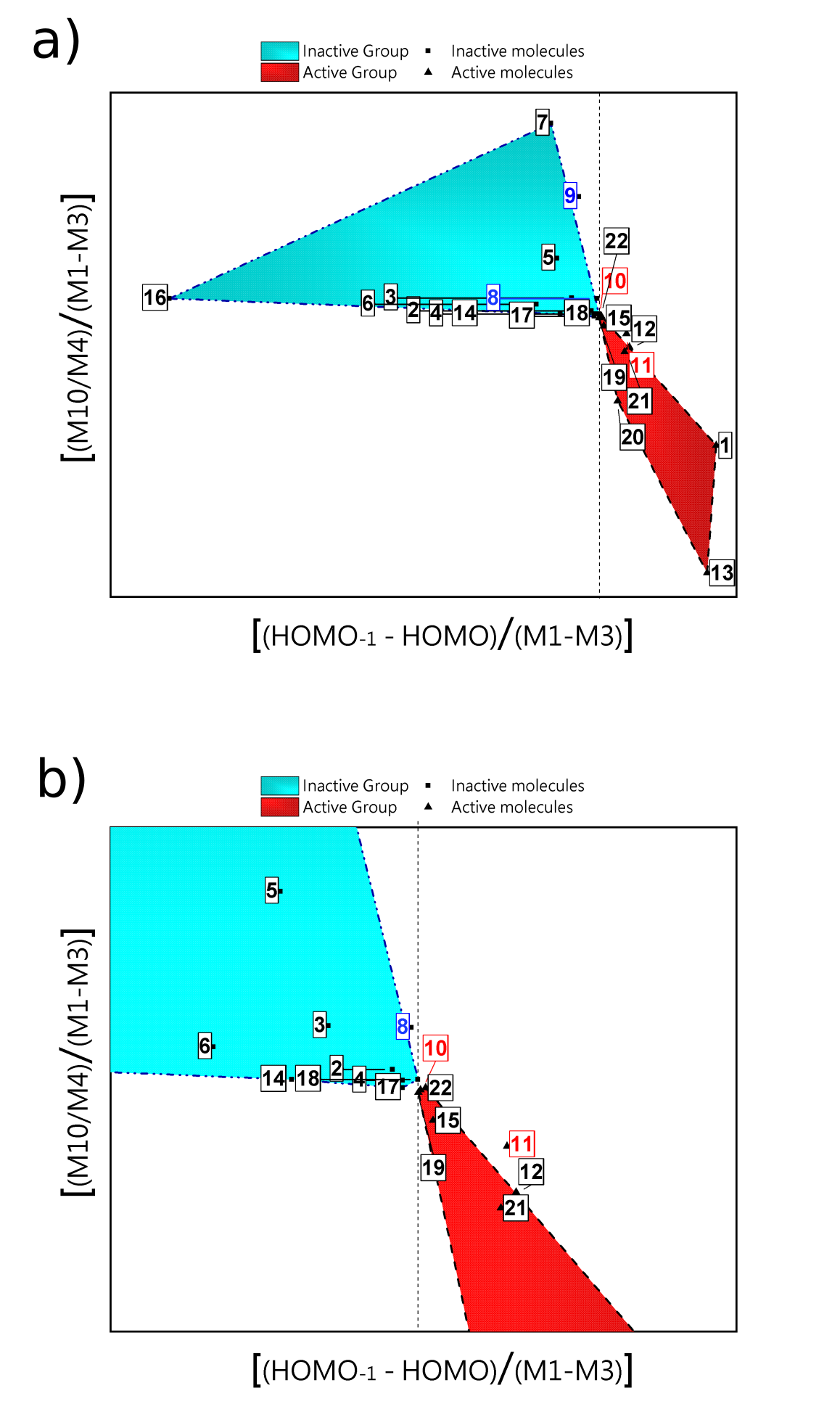}
  \caption{(a) A two-dimensional map of machine learning descriptors, including the test set molecules (8, 9, 10, and 11), calculated separately, and (b) a selected zoomed region of the map. The dashed line clearly indicates a group separation, primarily observed with the horizontal axis descriptor. The molecules previously discarded are color-highlighted.}
  \label{fgr:map-descriptors-3}
\end{figure}

In Figure \ref{fgr:map-descriptors-3}, we present the map of descriptors obtained through our reimplementation of machine learning using the combined dataset, consisting of both the training set and the test set. Notably, the fundamental characteristics of the model remain unchanged, but we observe a slight change in the boundary definitions. More specifically, when considering the model with all entries, the intersection of the groups shifts from -3.20 to -3.05. This shift indicates that there is no significant overfitting, suggesting that our model is not excessively reliant on specific data points.
Additionally, we find that the descriptor (HOMO$_{-1}$-HOMO)/(M1-M3) continues to effectively be able to distinguish between the two distinct groups of molecules: active and inactive. This is evident from the horizontal division observed in the inset of Figure \ref{fgr:map-descriptors-3}.

The physical interpretation of the factor $\Delta H$ in the context of structure-activity relationships still remains somewhat elusive, despite its frequent use in studies involving the IC$_{50}$ index and other biological quantifiers \cite{barone2000electronic, braga2004benzo, coluci2002identifying, troche2005carcinogenic, braga2003structure}. The HOMO energy, commonly associated with ionization potential and nucleophilic regions indicative of enhanced reactivity, is significantly important. This is further supported by the presence of a charge difference within the descriptor (M1-M3), precisely located in highly reactive regions, in general, associated with reactive bond sites.

\begin{figure}[p]
\centering
  \includegraphics[height=18cm]{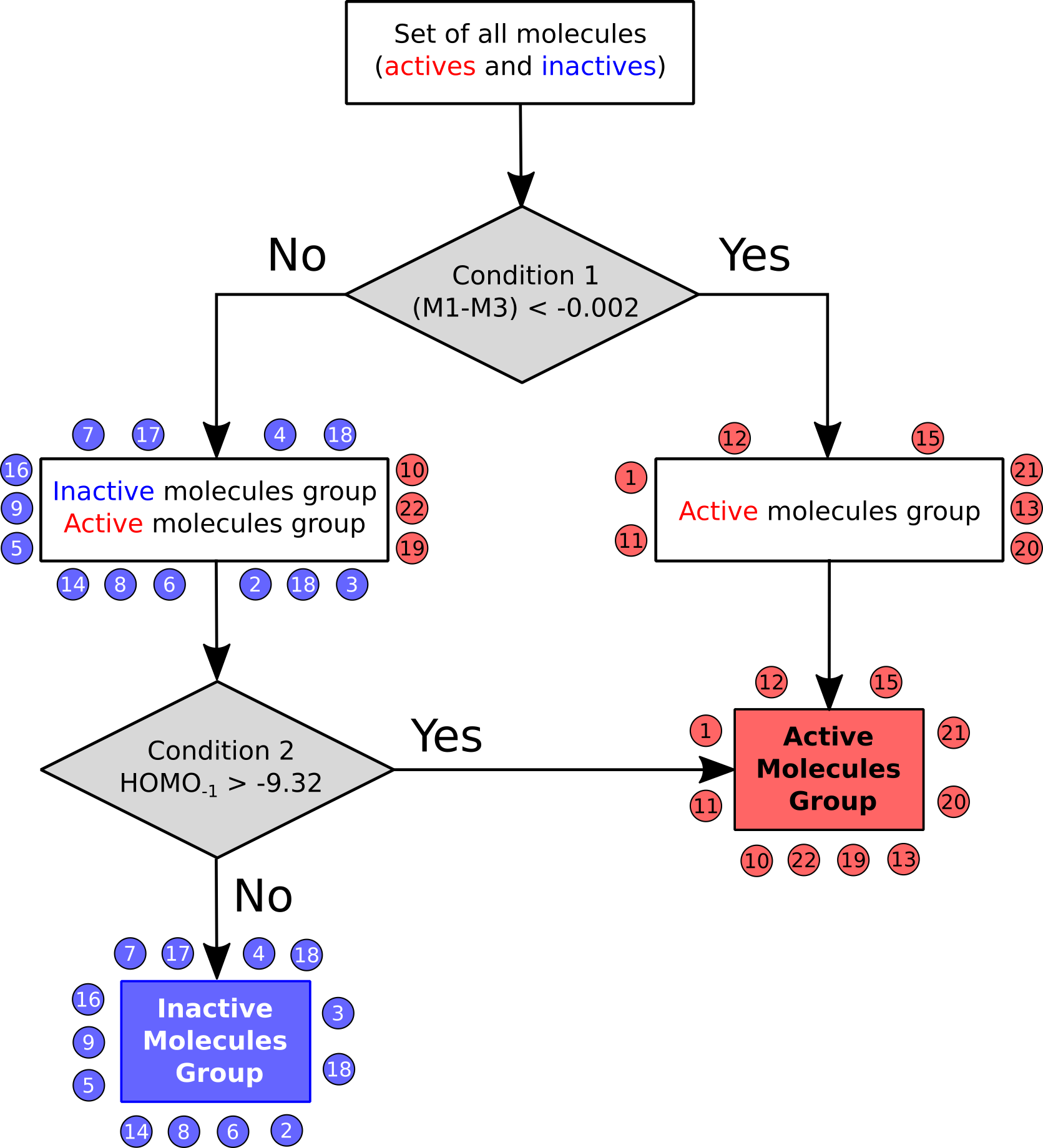}
  \caption{Simplified Boolean tree diagram. The diagram illustrates that two conditions are adequate to entirely segregate the two classes of molecules.}
  \label{fgr:boolean-tree}
\end{figure}

Notably, the ML horizontal descriptor alone can with which accuracy classify the active/inactive molecules. This validates that a simple rule is able to establish the relationship between molecular structure and biological activity. To further validate this observation, we build a Boolean decision tree (Figure \ref{fgr:boolean-tree}). By applying the following two rules, we achieved a 100\% accuracy in the separation of inactive and active groups:

\begin{itemize}
    \item Condition 1: The charge difference between atoms 1 and 3 (M1 and M3) is less than -0.002.
    \item Condition 2: The energy value of the penultimate occupied molecular orbital (HOMO$_{-1}$) is higher than -9.32.
\end{itemize}

According to condition 1, we were able to successfully differentiate all molecules except for numbers 10, 19, and 22, which were mistakenly categorized as inactive. However, applying the second rule, we rectified this misclassification by reassigning these molecules to the appropriate active group.

\section{Conclusions}
In summary, the difference in charges (M1-M3) observed in both descriptors indicates the relevance of the reactivity in the aromatic region between atoms 1 and 3 in influencing the biological behavior of these molecules. Moreover, the horizontal descriptor proves adequate in determining group behavior, while the difference between (HOMO$_{-1}$-HOMO) energy values was also identified as a crucial parameter for our model. Although we cannot assert with absolute certainty that our model is definitive, owing to limitations in sample size, the results remain consistent with other studies in the field, further underscoring the pivotal role of $\Delta H$ in structure-activity correlations.
Our study presents a streamlined methodology for optimizing the search process for molecules with specific biological activity, exemplified here for the chalcone skeleton. By using common skeletal structures to design new molecules (for instance, incorporating different chemical groups), we can, in principle, predict the activity/inactivity (based on the IC$_{50}$ index) of these novel molecules through the parameter features identified by the ML models. Notably, these predictions rely solely on electronic structure factors, utilizing the [(HOMO${-1}$-HOMO)/(M1-M3)] descriptor calculated via the PM3 method based. The methodology proposed here is completely general and can used for other classes of compounds. Studies along these lines are in progress.

\begin{acknowledgement}

The authors would like to thank CNPq (Grants 308428/2022-6 and  150595/2023-9). The authors would also like to thank CCM-UFABC for the computational resources.

\end{acknowledgement}

\begin{suppinfo}

Figure \ref{S1} show the statistical analysis. In this figure, each pixel corresponds to the Pearson coefficient (CP), which represents the correlation between two features.

\begin{figure}[p]
    \centering
    \includegraphics[height=14.5cm]{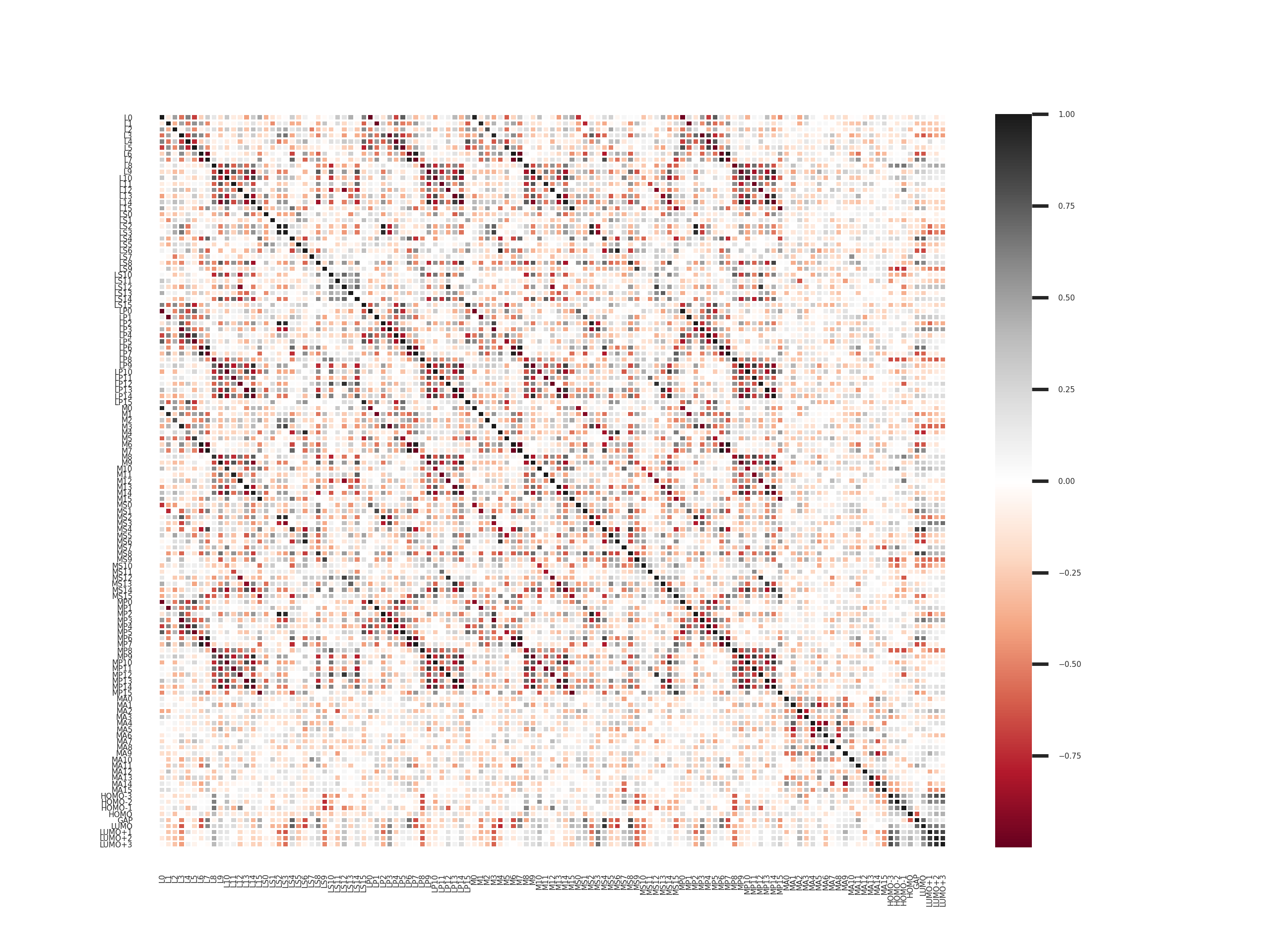}
    \caption{Pearson coefficient (CP). Each pixel represents the correlation between two features.}
    \label{S1}
\end{figure}

\end{suppinfo}

\bibliography{achemso-demo}

\end{document}